# A-site Cation disorder engineering in Ruddlesden-Popper Layered Perovskite Oxide La$_2$(Ba,Sr)In$_2$O$_7$ for Ferroelectricity

*Takumi Terauchi,[1] Wei Yi,[1,*] Rikuto Takada,[1] Hirofumi Akamatsu,[2] Ryo Ota,[3] Shuki Torii,[4] and Koji Fujita[1,*]*

[1]Department of Material Chemistry, Graduate School of Engineering, Kyoto University, Katsura Nishikyo-ku, Kyoto 615-8510, Japan

[2]Department of Applied Chemistry, School of Engineering, Kyushu University, Motooka, Fukuoka 812-0053, Japan

[3]HVEM Laboratory at Center for Advanced Research of Energy and Materials, Graduate School of Engineering, Hokkaido Uni versity, Sapporo 060-8628, Japan

[4]Institute of Materials Structure Science, High Energy Accelerator Research Organization (KEK), Ibaraki, 319-1106, Japan

ABSTRACT: The strategic design of ferroelectric materials exhibiting robust and reversible spontaneous polarization remains a pivotal challenge in functional materials research. Here, A-site cation disorder engineering is employed in the $n = 2$ Ruddlesden-Popper layered perovskite La$_2$Ba$_{1-x}$Sr$_x$In$_2$O$_7$ to achieve room-temperature ferroelectricity. Systematic substitution of Sr$^{2+}$ for Ba$^{2+}$ drives symmetry transitions from a parent centrosymmetric (CS) $P4_2/mnm$ structure ($x = 0$) to two emergent phases: a CS *Amam* phase ($0.3 \leq x \leq 0.4$) and a polar $A2_1am$ phase ($0.5 \leq x \leq 0.9$). Multimodal characterization combining synchrotron diffraction, neutron scattering, nonlinear optical spectroscopy, and hysteresis loop of electric polarization versus electric field reveals a hybrid improper ferroelectric (HIF) mechanism in the $A2_1am$ phase, arising from trilinear coupling between octahedral rotations and tilts. Cation disorder at A-sites suppresses the interfacial rumpling-induced octahedral elongation (deformation) while enhancing the octahedral rotations which are critical for the polar symmetry stabilization. First-principles calculations further elucidate that Sr/La disorder mitigates electrostatic interactions, enabling oxygen octahedral distortions necessary for ferroelectricity. This work establishes cation disorder engineering as a versatile strategy to design high-temperature multiferroics in layered perovskites, advancing the coupling between structural distortions and functional responses in complex oxides.

## 1. INTRODUCTION

Ferroelectric material, which possesses switchable spontaneous electric polarizations, is of great significance in the field of materials science and industry.[1-4] Among various structural families, perovskite oxides and their derivatives have emerged as particularly promising candidates due to their chemical versatility and structural adaptability. This versatility has enabled multiple design strategies for creating noncentrosymmetric polar materials across simple $ABO_3$ perovskite, double perovskites, and layered perovskite variants. Notably, quasi-two-dimensional Ruddlesden-Popper (PR)[5,6] and Dion-Jacobson (DJ)[7,8] phases have garnered significant attention following the recently discovered hybrid improper ferroelectricity (HIF), a mechanism enabling polarization through trilinear coupling of nonpolar structure distortions.[9,10]

The proper mechanism of the perovskite-type ferroelectrics, like $ABO_3$-type simple oxides with off-center cations, relies on the second-order Jahn-Teller (SOJT) effect,[11-13] in which polarization arises from covalent interactions between $nd^0/6s^2$ cations (e.g., Ti$^{4+}$, Bi$^{3+}$) and oxygen ligands.[14-16] However, this mechanism imposes strict compositional constraints, as many SOJT-active systems fail to achieve ferroelectric ordering.[17-19] Double perovskites ($AA'BB'O_6$) offer alternative pathways through charge-ordered layered structures, yet high-temperature synthesis often favors disordered cation configurations due to entropic stabilization.[29-31] The HIF paradigm revolutionizes this landscape by decoupling polarization from specific cation electronic configurations, instead utilizing cooperative oxygen octahedral rotations (OOR) and tilts (OOT) to break inversion symmetry.[32-45]

RP-phase layered perovskites ($A_{n+1}B_nO_{3n+1}$) generally consists of alternately stacked ($ABO_3$)$_n$ perovskite-slabs (where $n$ is the number of simple perovskite $ABO_3$ layer) and ($AO$) rocksalt-layers, presenting unique opportunities for symmetry engineering.[5,6] In this sandwiched structure, the absence of three-dimensional symmetry protection opens up the possibility of inversion symmetry breaking through the octahedral distortions. The $n = 2$ (double-layer) RP phase perovskites ($A_3B_2O_7$) has proven particularly fruitful for HIF realization due to the overall electrical polarization generated possiblly by the opposite but uncancelled in-plane dipoles (antiferroelectric).[46-50] This antiferroelectric ordering in the orthorhombic $A2_1am$ (#36) phase emerges from trilinear coupling of in-phase OOR and out-of-phase OOT.[51] This structural evolution is governed by the Goldschmidt tolerance factor $t$, which is defined as $t = (r_A + r_O)/\sqrt{2}(r_B + r_O)$, where $r_O$, $r_A$ and $r_B$ are Shannon's six-coordinate ionic (average) radii of $O^{2-}$ and cations on A- and B-sites, respectively.[52, 53] The $t < 0.89$ favoring the distortion modes essential for polarization is successful in $A^{2+}_3B^{4+}_2O_7$ systems, such as $(Ca,Sr)_3Ti_2O_7$ solid solution.[54] However, the implementation in hetero-valent $Ln^{3+}_2A^{2+}B^{3+}_2O_7$ compositions faces significant challenges due to enhanced interfacial rumpling at the perovskite-rocksalt boundaries.

As a general feature of all RP phase layered perovskite including the aristotype structure in tetragonal $I4/mmm$ (#139) space group, this interfacial rumpling arises from the symmetry breaking along the stacking direction and accompanies with the elongational deformation of adjacent octahedra along the stacking direction. Compared to the RP phase $A^{2+}_3B^{4+}_2O_7$ with equivalent A-cations, the rumpling in RP phase $Ln^{3+}_2A^{2+}B^{3+}_2O_7$ ($Ln$ = lanthanides) with hetero-valent A-cations ($Ln^{3+}$ and $A^{2+}$) is remarkable, due to the positively charged $[LnO]^+$ rocksalt layer attracting the negatively charged $O^{2-}$ ions at the adjacent rocksalt-perovskite interfaces. The enhanced rumpling magnitude gives rise to an overall suppression of OOR and results in the scarcity of RP phase $Ln_2AB_2O_7$ in $A2_1am$ structure even as $t < 0.89$.[55-57] Most of them have nonpolar tetragonal $P4_2/mnm$ (#136) centrosymmetric (CS) structure with one OOT, such as $La_2BaIn_2O_7$ ($t = 0.816$)[58, 59] or nonpolar orthorhombic $Fmmm$ (#69) CS-structure without any OOR or OOT, such as $La_2SrSc_2O_7$ ($t = 0.873$).[60] Recently, it has been reported that distorted structure of ferroelectric $La_2SrSc_2O_7$ is stabilized via the disordered A-cations $Sr^{2+}$ and $La^{3+}$, which suppresses the rumpling-induced octahedral deformations in competition and restores the OOR required for the HIF mechanism.[61]

In this context, $n = 2$ RP phase $La_2Ba_{1-x}Sr_xIn_2O_7$ (LBSI-$x$) system is of great interest as a model for HIF engineering in hetero-valent PR phase. This system combines three critical design elements: (1) equivalent cation substitution ($Ba^{2+}$ substitution with $Sr^{2+}$) at A-site to modulate tolerance factor while maintaining charge balance; (2) strategic introduction of cations disorder ($Sr^{2+}$ and $La^{3+}$) to mitigate interfacial rumpling; and (3) complete absence of SOJT-active cations ($In^{3+}$: $4d^{10}$) to eliminate SOJT-driven polarization, isolating HIF as the sole mechanism. Here, we establish a temperature-composition (T - $x$) diagram of the LBSI-$x$ solid solution for $0.0 \leq x \leq 0.9$ through comprehensive synthesis and characterization, revealing remarkable ferroelectric functionality.

Structural analyses and ferroelectricity studies of this LBSI-$x$ series were systematically conducted using a combination of synchrotron X-ray diffraction (SXRD), neutron powder diffraction (NPD), optical second harmonic generation (SHG), and electric polarization versus electric field (*P-E*) hysteresis loop. The atomic A-site cations distributions were examined by the high-angle annular dark-field scanning transmission electron microscopy (HAADF-STEM) observations and energy dispersive X-ray spectroscopy (EDS). The role played by the chemical substitution-induced two non-polar structure distortions, out-of-phase OOT and in-phase OOR, in stabilizing ferroelectricity is determined by combination of experiments results and first-principles density functional theory (DFT) calculations of two systems of $La_2BaIn_2O_7$ and $La_2SrIn_2O_7$.

Our results demonstrate that A-site disorder engineering suppresses interfacial rumpling, renormalizes OOR, and stabilizes switchable polarization up to 1000 K. This work not only advances fundamental understanding of structure-property relationships in HIF systems but also establishes cation disorder engineering as a powerful strategy for designing novel ferroelectrics beyond conventional SOJT paradigms. The demonstrated approach opens new avenues for developing functional materials in energy storage, piezoelectric transduction, and multiferroic applications.

## 2. RESULTS

### A. Room-Temperature Ferroelectric Structures

The NPD pattern of $La_2BaIn_2O_7$, the end member of $La_2Ba_{1-x}Sr_xIn_2O_7$ (LBSI-$x$) series for $x = 0.0$, is matched well with the reported tetragonal structure model, space group $P4_2/mnm$. It is observed that the tetragonal lattice remains until 20 mol% of Sr substitution ($x = 0.2$). Beyond $x = 0.2$, the splitting of characteristic reflections (e.g., 202 and 022 shown in Figure 1a) imply an orthorhombic phase, which continues until $x = 0.9$. The other end member of LBSI-$x$ series for $x = 1.0$ (full substitution of Sr for Ba), $La_2SrIn_2O_7$ could not be synthesized due to phase separation. Based on the orthorhombic unit cell, all the observed reflection conditions of LBSI-$x$ ($0.4 \leq x \leq 0.9$) are $hkl$: $k + l = 2n$, $0kl$: $k + l = 2n$, $h0l$: $h, l = 2n$, $hk0$: $k = 2n$, $h00$: $h = 2n$, $0k0$: $k = 2n$, and $00l$: $l = 2n$ ($n$: integer). Two space groups, polar $A2_1am$ ($Cmc2_1$ in standard setting) and nonpolar $Amam$ ($Cmcm$ in standard setting), can be derived. For $x = 0.3$, the appearance of 118 reflection in $P4_2/mnm$ phase as well as the observed 109 reflection and splitting reflections (202 and 022) in orthorhombic $Amam$ or $A2_1am$ phases suggest the coexistence of orthorhombic and tetragonal phases (Figure 1a).

To distinguish the acentric nature of LBSI-$x$ series, we measured optical SHG, which is one of the powerful tools to probe inversion symmetry breaking in piezoelectrics, pyroelectrics, and ferroelectrics. Figure 1b displays the SHG intensities of LBSI-$x$ for $0 \leq x \leq 0.9$ at room temperature. The inset of Figure 1b shows the obvious distinction between the inactivated and activated SHG results for $x = 0.2$ and 0.7, respectively. The significant SHG-activated signals, indicative nature of NCS structures, are clearly observed for $0.5 \leq x \leq 0.9$, allowing us to unambiguously conclude the polar $A2_1am$ phase at room-temperature. The CS and nonpolar $Amam$ phase for the orthorhombic LBSI-$x$

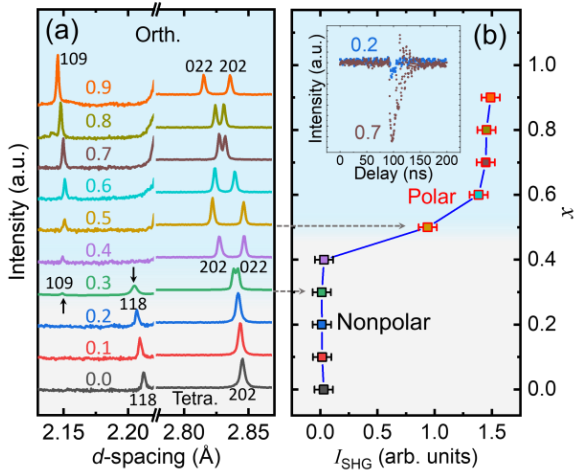

Figure 1. (a) Enlarged NPD patterns of LBSI-$x$ ($0.0 < x < 0.9$) at selected $d$-spacing ranges, in which the 202 reflections shift towards small $d$ and splits into two reflections, 202 and 022, as the Sr solid solubility increases. (b) SHG integral intensity of LBSI-$x$ ($0.0 < x < 0.9$) at room temperature. The inset shows the inactivity and activity of LBSI-0.2 and -0.7, respectively.

($0.3 \leq x \leq 0.4$) are verified simultaneously by the inactivated SHG.

The lattice parameters and unit cell volumes as a function of Sr solid solution ($x$) are obtained by the Le Bail-method refinements against the NPD data at room temperature (Figure 2). With the increase in the substitution of relatively smaller Sr, the almost continuous linear shrinkage of the unit cell volume is observed across the range $0.0 \leq x \leq 0.9$. The non-monotonical variation of lattice parameters and $a/b$ ratio verifies the occurred phase transition. At $x = 0.3$, the discontinuous change of lattice parameters $a$ and $b$ originates from the tetragonal-to-orthorhombic phase transition which also result in the observed splitting reflections (Figure 1a). The two-phases coexistence of $Amam$ and $P4_2/mnm$ around $x = 0.3$ are verified by the better fitting result (weighted profile $R$-factor, $R_{wp}$ = 4.31%), compared to that of individual $Amam$ ($R_{wp}$ = 8.45%) or $P4_2/mnm$ ($R_{wp}$ =14.88%). In the orthorhombic phase region, the evolution of lattice parameters with $x$ shows a well-defined minimum and gradient change in the $a/b$ ratio at $x = 0.5$. Combined with the studied SHG, this suggests that a composition-driven structural transition occurs between $Amam$ and $A2_1am$ phases. In the case of reported polar $n = 2$ RP ($Ca_ySr_{1-y}$)$_{1.15}$Tb$_{1.85}$Fe$_2$O$_7$ - Ca$_3$Ti$_2$O$_7$ solid solution, this OOR-induced gradient change of in-plane lattice parameters ratio has been observed also.[32] The polymorphic phase diagram of LBSI-$x$ solid solution at room temperature, as shown in Figure 2, is thus established by the diffraction studies combined with SHG.

We carried out ferroelectric measurements for $x = 0.7$ at room temperature using the remanent hysteresis method. The well-shaped rectangular $P-E$ hysteresis loops indicate the ferroelectricity with switchable polarization (see Figure S1). The net remanent polarization and coercive electric field are around 0.023 $\mu$C/cm$^2$ and 150 kV/cm, respectively. To our knowledge, this is the first observation of ferroelectricity in In-based perovskite without SOJT-active cations (with $ns^2$ or $nd^0$).

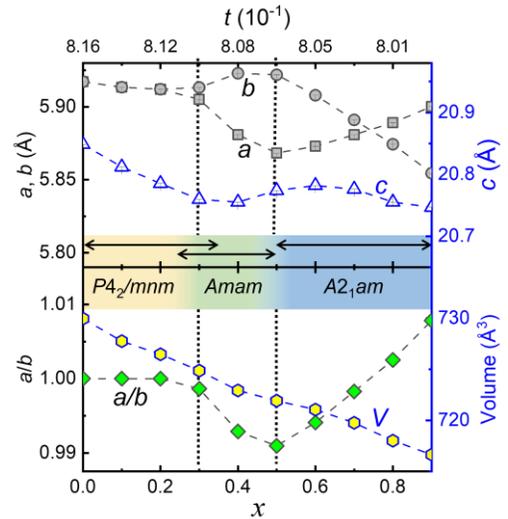

Figure 2. The Sr solid solubility ($x$) dependence of lattice parameters $a$ and $b$ (upper-left y-axis), $c$ lattice parameter (upper-right y-axis) the ratio $a/b$ (down-left y-axis) and volume of unit cell (V) (down-right y-axis) for LBSI-$x$. The tolerance factor $t$ for LBSI-$x$ is shown from the top axis. The room-temperature phase diagram of LBSI-$x$ is established in the present study. factor $t$ for LBSI-$x$ is shown from the top axis. The room-temperature phase diagram of LBSI-$x$ is established in the present study.

### B. Disordered A-site cations

In the layered structure of $n = 2$ RP phase $AO(ABO_3)_2$, there are two distinct A-sites, a 12-coordinated A1-site in perovskite slab ($A_{Pv}$-site) and a 9-coordinated A2-site in rocksalt layer ($A_{Rs}$-site). The distribution of A-site cations (Ba, Sr, and La) over the $A_{Pv}$- and $A_{Rs}$-sites in LBSI-$x$ ($x = 0.0, 0.4$ and $0.7$) was examined by the atomic scale STEM-EDS mapping, which provides insight into the layer stacking and spatially resolved distribution of cations. Figure 3a and 3c display the high-angle annular dark-field scanning transmission electron microscopy (HAADF-STEM) images with the inset of crystal structures in the [11$\bar{2}$0] projection for $x$ = 0.4 and 0.7, respectively. The intensity is expected to scale approximately as the square of the atomic number Z, resulting in the O columns ($Z_O$ = 8) are invisible compared to the bright spots of have cation columns ($Z_{Sr}$ = 38, $Z_{Ba}$ = 56, $Z_{La}$ = 57, and $Z_{In}$ = 49). From the projected crystal structure shown with the insetting unit cell structures, it is noticed that the $A_{Pv}$- and $A_{Rs}$-site columns form alternate layers along the [001] direction. In the combined elemental map with Sr, In, Ba, and La atoms (see Figure 3b and 3d), the color spots represent the distinct elements identified by EDS. The In-columns show the pure green spots, and the $A_{Pv}$- and $A_{Rs}$-site columns show the orange and purple spots, respectively, owing to the mixing of red (Ba), yellow (Sr) and blue (La). The net intensity profiles of distinct cations for $x$ = 0.4 (inset of Figure 3b) and 0.7 (inset of Figure 3d)

Table 1. Selected crystallographic data and structure parameters obtained by combined refinements against the SXRD and NPD data of the LBSI-0.2, 0.4, and 0.7 at room temperature. $g$: occupancy parameter; $U_{aniso}$: anisotropic thermal parameter; GOF: goodness of fitting. [a] Refined isotropically.

| Atom | site | g | x | y | z | $U_{iso}$ (Å²) |
|---|---|---|---|---|---|---|
| Ba1 | 4f | 0.8 | 0.25860(8) | x | 0 | 0.01057(12) |
| Sr1 | | 0.0561(6) | | | | |
| La1 | | 0.1439(6) | | | | |
| Sr2 | 8j | 0.0719(3) | 0.27108(4) | x | 0.184471(17) | 0.00814(9) |
| La2 | | 0.9280(3) | | | | |
| In1 | 8j | 1.0 | 0.25570(14) | x | 0.39993(4) | 0.0050(2) |
| O1 | 4g | 1.0 | 0.29733(13) | x | 0.5 | 0.0338(3) |
| O2 | 8j | 1.0 | 0.18954(7) | x | 0.29103(3) | 0.0188(2) |
| O3 | 8h | 1.0 | 0.0 | 0.5 | 0.39004(5) | 0.0188(3) |
| O4 | 4e | 1.0 | 0.5 | 0.5 | 0.37014(5) | 0.0093(3) |
| O5 | 4e | 1.0 | 0.0 | 0.0 | 0.41103(6) | 0.0150(3) |

La$_2$Ba$_{0.8}$Sr$_{0.2}$In$_2$O$_7$: Space group: $P4_2/mnm$ (#136), Z = 4. Lattice parameters: $a$ = $b$ = 5.913549(9) Å, $c$ = 20.78775(5) Å, V = 726.949(2) Å³. Reliability indices: $R_p$ = 4.33% and $R_{wp}$ = 5.52%, GOF = 2.43.

| Atom | site | g | x | y | z | $U_{iso}$ (Å²) |
|---|---|---|---|---|---|---|
| Ba1 | 4c | 0.6 | 0.25 | 0.23567(7) | 0 | 0.0130(1) |
| Sr1 | | 0.2042(3) | | | | |
| La1 | | 0.1958(3) | | | | |
| Sr2 | 8g | 0.0979(14) | 0.25 | 0.22810(5) | 0.18497(1) | 0.0096(1) |
| La2 | | 0.9021(14) | | | | |
| In1 | 8g | 1.0 | 0.25 | 0.24140(18) | 0.40020(4) | 0.005(1)[a] |
| O1 | 4c | 1.0 | 0.25 | 0.17531(19) | 0.5 | 0.0404(5) |
| O2 | 8g | 1.0 | 0.25 | 0.32310(11) | 0.29217(3) | 0.0431(3) |
| O3 | 8e | 1.0 | 0.0 | 0.5 | 0.37683(3) | 0.0189(2) |
| O4 | 8e | 1.0 | 0.5 | 0.5 | 0.40617(4) | 0.0297(3) |

La$_2$Ba$_{0.6}$Sr$_{0.4}$In$_2$O$_7$ Space group: $Amam$ (#63), Z = 4. Lattice parameters: $a$ = 5.88177(13) Å, $b$ = 5.92453(13) Å, $c$ = 20.7577(5) Å, V = 723.34(3) Å³. Reliability indices: $R_p$ = 4.22% and $R_{wp}$ = 5.32%, and GOF = 2.39.

| Atom | site | g | x | y | z | $U_{iso}$ (Å²) |
|---|---|---|---|---|---|---|
| Ba1 | 4a | 0.3 | 0.20971(9) | 0.23260(12) | 0 | 0.0144(3) |
| Sr1 | | 0.3839(6) | | | | |
| La1 | | 0.3160(6) | | | | |
| Sr2 | 8b | 0.1580(3) | 0.27065(10) | 0.23419(9) | 0.18555(3) | 0.0123(2) |
| La2 | | 0.8420(3) | | | | |
| In1 | 8b | 1.0 | 0.25* | 0.2433(3) | 0.40117(6) | 0.0091(3)[a] |
| O1 | 4a | 1.0 | 0.2779(1) | 0.1558(3) | 0.5 | 0.0199(5) |
| O2 | 8b | 1.0 | 0.2221(1) | 0.3349(2) | 0.29386(5) | 0.0240(4) |
| O3 | 8b | 1.0 | 0.0207(2) | -0.0266(3) | 0.37551(4) | 0.0172(4) |
| O4 | 8b | 1.0 | 0.4631(2) | 0.5299(3) | 0.41447(5) | 0.0263(5) |

La$_2$Ba$_{0.3}$Sr$_{0.7}$In$_2$O$_7$ Space group: $A2_1mam$ (#36), Z = 4. Lattice parameters: $a$ = 5.88299(17) Å, $b$ = 5.89319(17) Å, $c$ = 20.789(6) Å, V = 720.750(1) Å³. Reliability indices: $R_p$ = 5.29% and $R_{wp}$ = 6.95%, and GOF = 3.11. *The $x$-value of In1 is fixed as an origin of the polar $a$-axis.

verify the disordered distributions of Sr/La at $A_{Rs}$-sites, and Ba/Sr/La at $A_{Pv}$-sites, respectively. The approximate background intensity of Ba at $A_{Rs}$-site indicates that Ba occupies only $A_{Pv}$-sites for all LBSI-$x$. Compared to that for $x = 0.4$, the noticeably lower intensity of $A_{Pv}$-sites Ba in $x = 0.7$ is due to its reduced solid solubility (increase of Sr). In addition, the intensity of $A_{Rs}$-site Sr increases while the intensity of $A_{Rs}$-site La decrease indicates that the increase in the solid solubility of Sr in LBSI-$x$ system simultaneously increases the occupancy of Sr at both $A_{Pv}$- and $A_{Rs}$-sites, thus enhancing the disorder of A-sites cations.

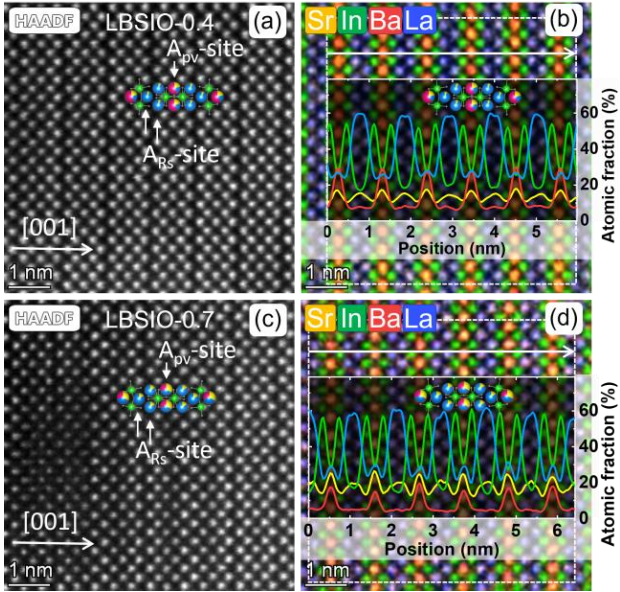

Figure 3. Atomically resolved HAADF image and crystal structure of LBSI-0.4 (a) and LBSI-0.7 (b) in [11$\bar{2}$0] projection. Combined EDS map of LBSI-0.4 (b) and LBSI-0.7 (d) for Sr (yellow), In (Green), Ba (red), and La (blue) elements. Atomic fraction intensity profiles in the marked areas by frames of dotted white lines in (b) and (d).

In further structure analysis, the combined Rietveld refinements were performed on all the compositions in the series against NPD (collected by three detection banks) and SXRD patterns at room temperature. The tetragonal $P4_2/mnm$ model, orthorhombic $Amam$ model, and orthorhombic $A2_1am$ model were used for the structure refinements of LBSI-$x$ for $0 \leq x \leq 0.2$, $x = 0.4$, and $0.5 \leq x \leq 0.9$, respectively and provide good fitting results. For $x = 0.3$, the good fitting result was obtained for the refinement with two-phase coexisting models of $P4_2/mnm$ (~63 vol.%) and $Amam$ (~37 vol.%) confirming the coexistence of the tetragonal and orthorhombic phases. Because of the negligible deviation in the stoichiometry, the refinements were performed with the constraints that the overall stoichiometry was maintained and that each site was fully occupied. The Sr/La distribution over the $A_{Pv}$-site and $A_{Rs}$- site are variable within the constraints involving that each site is fully occupied, that Ba occupies $A_{Pv}$-sites only, and that the overall nominal ratio of $Ba_{1-x}Sr_x$: La = 1: 2. This stoichiometric composition model provides good fitting to the NPD and SXRD data (see Figure S2 and Table S1, SI). Figure 4 shows the Rietveld refinement against the NPD patterns (from the backscattering bank) of representative LBSI-$x$ for $x = 0.2$, 0.4, and 0.7. The structural parameters obtained from the combined refinements against NPD and SXRD patterns for $x = 0.2$, 0.4, and 0.7 are listed in Table 1. Note that the occupancy ($g$) of Sr, Ba, and La as A site

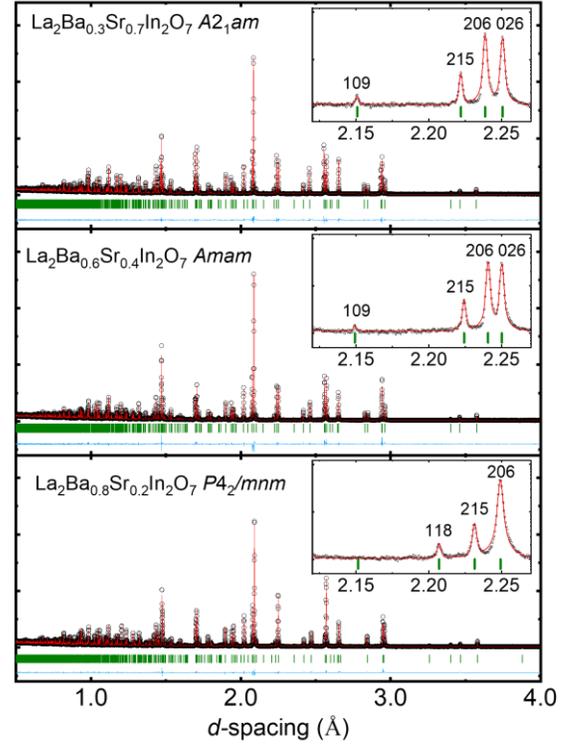

Figure 4. Observed, calculated, and difference plots from the structural refinement of LSI-0.2, 0.4, and 0.7 against NPD data at room temperature. Green tick marks indicate reflections for the majority phase.

cations is estimated by the diffraction studies combined with STEM-DES.

## C. Phase Transition at High Temperature

To identify the phase transition and crystal symmetry at high temperature, we analyzed the SXRD patterns of LBSI-$x$ for $0 \leq x \leq 0.9$ in temperature range of 300-1100 K and investigated their SHG activation in temperature range of 300-750 K. The phase diagram is shown in Figure 5. For $0 \leq x \leq 0.3$, which has a nonpolar tetragonal $P4_2/mnm$ phase at room temperature, the splitting reflections, like 206 and 026, indicate that the structure transitions from a tetragonal phase to an orthorhombic phase as the temperature increases. The structure transition temperature between two phases ($T_S$), which is estimated via the SXRD pattern with observable splitting reflections and the temperature dependence of lattice parameters (Figure S3, SI), gradually decreases from 900 K to 500 K with the solid solubility of Sr increasing from 0% to 30% (Figure S4, SI). Given all the observed reflection conditions of SXRD at high temperature, the possible settings of space groups, NCS $A2_1am$ and CS $Amam$, are suggested. The inactive SHG observed in temperature range of 300-750 K for both $x = 0.3$ ($T_S = 500$K) and 0.2 ($T_S = 700$ K) suggests that the tetragonal $P4_2/mnm$ phase of LBSI-$x$ for $0 \leq x \leq 0.3$ at RT transitions into an orthorhombic $Amam$ phase at high temperature.

For $0.4 \leq x \leq 0.9$, the reflection conditions of SXRD patterns at 1100 K are the same as those for the room-temperature phase, suggesting the possible settings of space groups, NCS $A2_1am$ and CS $Amam$. For $x = 0.4$, which has a nonpolar orthorhombic $Amam$ phase at room temperature and inactivated SHG up to 750 K, the nonpolar $Amam$ phase is assigned in temperature range of 300-1100 K. The ferroelectric-to-paraelectric phase transition temperatures for $x = 0.5$ and 0.6 were determined by the SHG intensity gradually decreasing and reaching near zero at (Curie Temperature) $T_C \sim 600$ K and $\sim 710$ K, respectively. In addition, the evolution of lattice parameters with temperature shows the well-defined minimum and gradient change in the $a/b$ ration at $T_C$ (see the inset of Figure 5). This result can be attributed to the OOR-reduced difference between lattice parameters $a$ and $b$.[32] Given the SHG activation for $x = 0.7$ persists up to 750 K, the ferroelectric-to-paraelectric phase transition temperature $T_C = 1000$ K for $x = 0.7$ is estimated by the minimum and gradient change in the $a/b$ ratio. The ferroelectric $A2_1am$ structure for $x = 0.8$ and 0.9 persists above 1100 K, since neither change in the $a/b$ ratio gradient nor inactivated SHG were observed in the temperature range studied.

## 3. DISCUSSION

### A. HIF Distortions with Disordered Sr/La.

The arisotype structure of $n = 2$ RP phase perovskite oxides adopts the nonpolar space group $I4/mmm$ (Figure 6a). Starting from the undistorted $I4/mmm$ aristotype phase, the experimentally observed three phases of tetragonal $P4_2/mnm$, orthorhombic $Amam$, and orthorhombic $A2_1am$ symmetries, can be explored by condensing individually or jointly various distortions (including OOR and OOT) accompanied by the enlarged unit cell metric, $\sqrt{2}a_p \times \sqrt{2}a_p \times c_p$ ($a_p$ and $c_p$ are the pseudo-tetragonal lattice parameters), as shown in Figure 6b. The octahedral distortions in nonpolar tetragonal $P4_2/mnm$ symmetry are noted as $(a^-b^0c^0)/(b^0a^-c^0)$ in Glazer notion, where the $a^-$ signifies a conventional single out-of-phase OOT about the basal plane axis of the aristotype $I4/mmm$ $(a^0a^0c^0)/(a^0a^0c^0)$.[62] The structural transformation like a nonpolar zone-boundary mode, known as the irreducible representation (irrep) $X_3^-(a;a)$, which stabilizes an out-of-plane antipolar distortion mode ($M_2^+$). The $X_3^-(a;0)$ mode for the symmetry lowing from the $I4/mmm$ to nonpolar $Amam$ $(a^-a^-c^0)/(a^-a^-c^0)$ is signified as $a^-a^-$, i.e. the out-of-phase OOT of equal magnitude about orthogonal axes. For the second-order phase transition from nonpolar $Amam$ to the polar $A2_1am$ $(a^-a^-c^+)/(a^-a^-c^+)$, the nonpolar distortion of $a^-a^-$ OOT couples with an additional in-phase OOR of different magnitude about the stacking axis, which is signified with $c^+$, corresponding to the nonpolar zone-boundary mode that transforms like the irrep $X_2^+$.

In HIF $A2_1am$ structure, these two nonpolar distortions of OOT and OOR are trilinearly coupled with the in-plane polar motions of ions, which transforms like the polar zone-center mode $\Gamma_5^-$. The layer-by-layer contribution of the polar displacements of cations and oxide ions along the orthorhombic $a$-axis [100] for LBSI-0.7 is depicted in Figure 6b. The opposite, but not cancelling, dipoles yielded by the relative displacements of cations and oxide ions in each layer generate overall electrical polarization, which are responsible for the ferroelectricity. In addition to the octahedral distortions above, the out-of-plane opposite motions of oxygens and cations at each interface naturally appear in RP systems due to symmetry breaking along the direction of layer stacking. The intensification of this interlayer rumpling ($\Gamma_1^+$), which is accompanied by the octahedral deformation, mainly originates from the rocksalt layer in RP $Ln^{3+}{}_2AB_2O_7$, such as $La_2BaIn_2O_7$ in which the $La^{3+}$ on $A_{Rs}$-sites produces the positive-charge electrostatics of the $[LaO]^+$ layer (Figure 6). In the presence of a large rumpling, the rotations more easily produce over-electrostatic attraction with negatively charged $O^{2-}$ ions. It consists with the suppression of rumpling for the ferroelectric $La_2SrSc_2O_7$ via A-site cations disorder.[61]

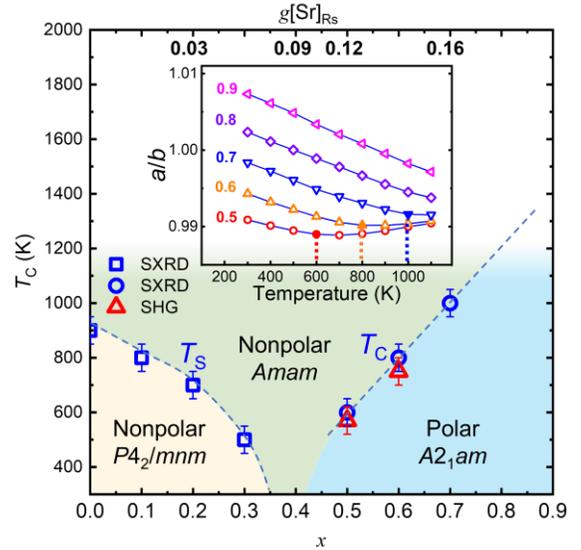

Figure 5 Structure phase diagram showing dependence of crystal structure on composition and temperature in $La_2Ba_{1-x}Sr_xIn_2O_7$ ($0 \leq x \leq 0.9$) series. Blue squares, $T_S$ from SXRD patterns; blue circles, $T_C$ from SXRD patterns; red triangles, $T_C$ from SHG intensities. Inset shows the Sr solid solubility ($x$) dependence of the a/b ratio for LBSI-$x$ ($0.5 \leq x \leq 0.9$) with the solid symbols indicating the paraelectric-to-ferroelectric phase transition temperature.

In RP perovskites with a small tolerance factor, the instability of octahedral distortions (OOT and OOR) and rumpling-induced octahedral deformations relates to their ability to decrease energy by optimizing A-cations and $O^{2-}$ coordination of $La^{3+}$ cations. The stronger competition of rumpling with OOR rather than OOT prevents appearance of OOR.[55]

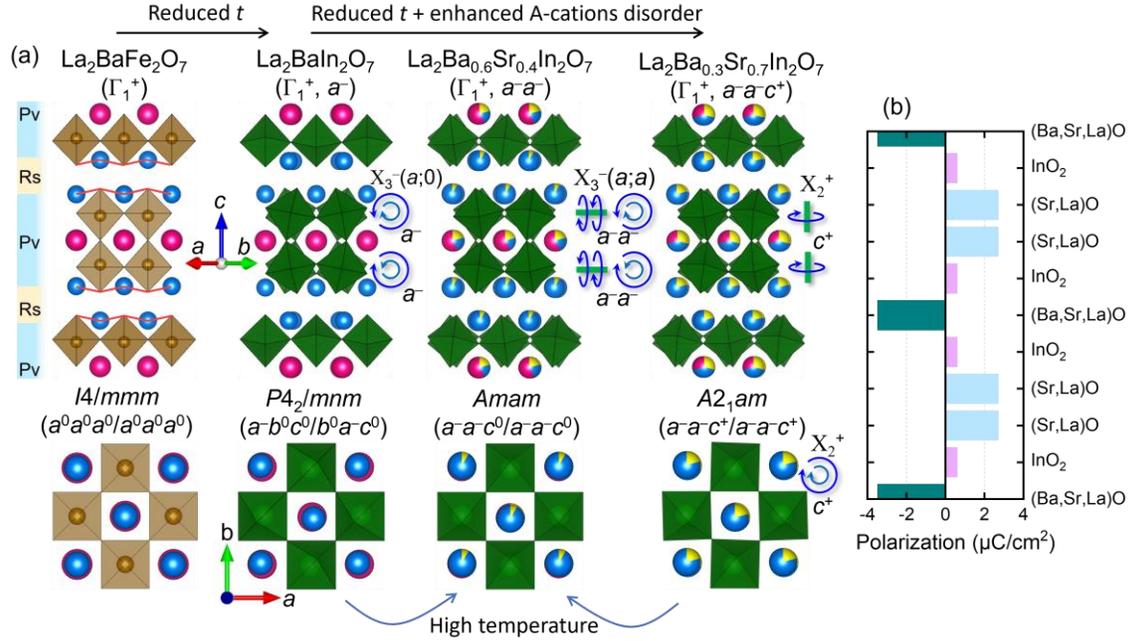

Figure 6. (a) Crystal structure side views (up) and top views (down) of space group symmetries in $n = 2$ RP structures consisting with alternately stacked perovskite layer and rocksalt layer. The blue, red, yellow, brown and green spheres denote the La, Ba, Sr, Fe and In, respectively. The brown and green quadrilaterals denote $FeO_6$ and $InO_6$ octahedra, respectively. (b) The relative displacements of cations and oxygens associated with polar distortion $\Gamma_5^-$ mode for $x = 0.7$. The magnitudes and directions of in-plane displacements of ions are presented by the horizontal length of solid bars.

The chemical substitutions-induced variation in the amplitudes of structural distortions for the present LBSI-$x$ series are examined (Figure 7). The undistorted $I4/mmm$ aristotype structures are used as references where the rumpling motions are artificially removed by putting A-site cations and oxide ions back to the closed-packed AO layer (see Table S2, SI). As the substitution of $Sr^{2+}$ for $Ba^{2+}$ in LBSI-$x$ series the rumpling $\Gamma_1^+$ is monotonically suppressed as well as the gradual increase of $Sr^{2+}$ occupancy on $A_{Rs}$-sites ($\alpha = g[Sr]_{Rs}$) accompanied by the corresponding decrease of $La^{3+}$ occupancy on $A_{Rs}$-sites (1- $\alpha$ = $g[La]_{Rs}$) which results in the enhancement of A-site cations disorder. The exchanging position of $La^{3+}$ and $Sr^{2+}$ is expected to suppress the rumpling due to the reduction of electrostatic field generated by the non-integer positive effective charge of $[Ln^{3+}{}_{1-\alpha}A^{2+}{}_\alpha O]^{<1-\alpha>+}$ rocksalt layers weakens the coordination. In $P4_2/mnm$ phase LBSI-$x$ series, the substitution of smaller $Sr^{2+}$ for $Ba^{2+}$ induces a slight increase in the amplitude of the one OOT, $X_3^-(a;a)$, as well as the negligible amplitude of $M_2^+$. The suppressed rumpling $\Gamma_1^+$, although not sufficient to restore the OOR, still induces a structural instability and caused the structural transition to $Amam$ phase with two OOT, $X_3^-(a;0)$, at room temperature (Figure 7c). The chemical substitution-induced $P4_2/mnm$-to-$Amam$ phase transition can be also observed in $Tb_2(Sr,Ca)Fe_2O_7$ solid solution at room temperature.[32**Error! Bookmark not defined.**] The structure transformation of $Amam$-to-$P4_2/mnm$ accompanied by the change in OOT sense occurs discontinuously (Figure 6b). Indeed, there is no group-subgroup relationship between $Amam$ and $P4_2/mnm$. The discontinuous nature of first-order phase transition is also suggested by the coexistence of two phases for $x = 0.3$ and the evolution of the lattice parameters ($a$ and $b$) with $x$ for LBSI-$x$ (Figure 2).

The chemical substitution-induced decrease of the $P4_2/mnm$-to-$Amam$ phase transition temperature, $T_S$, to room temperature for present LBSI-$x$ series (Figure 5) can be also attributed to the suppression of $\Gamma_1^+$ reduces the structural stability of $P4_2/mnm$. For the RP phase La$_2$SrSc$_2$O$_7$, it has been noted that the dependence of rumpling $\Gamma_1^+$ on temperature is minimal to negligible compared to the A-cations disorder.[61] It has been known that, in the $A_2B_2O_7$ RP-perovskites with negligible rumpling, such as (Ca,Sr)$_3$Ti$_2$O$_7$, the stabilization temperature of the $Amam$ phase is lower than that of $P4_2/mnm$.[63] The $Amam$ phase of LBSI-0.4 is unchanged on cooling to 100 K. the rumpling-induced suppression sequence of octahedral distortions is $X_2^+$, OOT, $X_3^-(a;0)$, OOT, $X_3^-(a;a)$. Due to the enhanced entropy induced by high temperatures, chemical substitution-induced smaller tolerance factor, and suppression of rumpling induced by A-cations disorder.

Within the orthorhombic phase LBIS-$x$ series, the substitution of Ba$^{2+}$ with Sr$^{2+}$ induces that a rise in the amplitudes of $X_3^-(a;0)$ is rapid in $Amam$ phase and becomes relatively flat after crossing the $Amam$-to-$A2_1am$ phase transition (Figure 7c). The substitution-induced reduction of both tolerance factor and rumpling $\Gamma_1^+$ lead to the rapid rise in the amplitude of $X_3^-(a;0)$ as an only way to stabilize the $Amam$ structure. The chemical substitution-induced $Amam$-to-$A2_1am$ phase transition is accompanied by the appearance of OOR $X_2^+$. The rise in amplitude of $X_3^-(a;0)$ becomes relatively flat after the substitution-induced phase transition due to the incremental amplitudes of $X_2^+$ as another way to stabilize the $A2_1am$ structure. The observable amplitudes of polarization $\Gamma_5^-$ in $A2_1am$ phase as well as the $T_C$ of $Amam$-to-$A2_1am$ phase transition increase with the introduce of Sr$^{2+}$ suggests that the A-cation disorder plays a role in stabilizing the HIF $A2_1am$ phase. Note that the further A-cation substitution in Tb$_2$(Sr,Ca)Fe$_2$O$_7$ could not produce the OOR $X_2^+$ until phase separation due to the equivalent substitution of smaller Ca$^{2+}$ for Sr$^{2+}$ on A$_{Pv}$-site rather than for La$^{3+}$ on A$_{Rs}$-sites only reduces the tolerance factor but barely inhibits the rumpling. Considering the absence of SOJT-active cations, the HIF mechanism is assigned to these LBSI-$x$ series in the $A2_1am$-distorted structure.

**B. Impact of Cation Disorder on Structural Instability**

To examine the impact of cation distributions on the structural instability of La$_2$(Ba,Sr)In$_2$O$_7$ solid solution, we performed first-principles calculations of La$_2$BaIn$_2$O$_7$ and La$_2$SrIn$_2$O$_7$ two individual systems with ordering and disordering A-cations distributions. First, the stable structure exploration was conducted by finding unstable phonon modes,[64] starting from a 2 × 2 × 1 supercell (48 atoms) with $I4/mmm$ space group symmetry, in which the Sr, Ba and La cations are placed on their expected sites, that is, Sr and Ba on the 12-coordinate A$_{Pv}$-site and La on the 9-coordinate A$_{Rs}$-site, respectively. In the ordered cations distribution systems, only zone center phonon modes of the supercell are traced because the $X$- and $M$-point zone boundary modes of the primitive $I4/mmm$ cell are folded into the $\Gamma$-point of the supercell. This exploration results in finding polymorphs with inequivalent structure more stable than the parent $I4/mmm$ structure. Three most stable inequivalent structures shown in (Figure S6, SI) includes the most stable $Pnam$ [$a^-a^-c^+/a^-a^-(-c)^+$] (#62) and $P4_2/mnm$ structure for La$_2$SrIn$_2$O$_7$ (LSIO) and La$_2$BaIn$_2$O$_7$ (LBIO), respectively, and the $A2_1am$ structure, which is the most stable polar structure with OOR distortion for both systems.

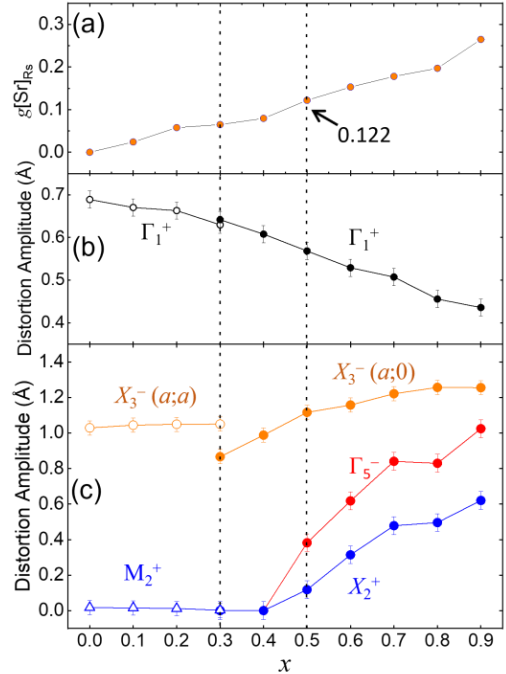

Figure 7. Evolutions of distortion modes (a) occupancy of Sr at A$_{Rs}$-site $g$[Sr]$_{Rs}$, (b) $\Gamma_1^+$ (rumpling), and (c) OOT $X_3^-(a;a)$ and $M_2^+$ (antipolar) in $P4_2/mnm$ phase, OOT $X_3^-(a;0)$ in $Amam$ phase, and OOT $X_3^-(a;0)$, OOR $X_2^+$, and $\Gamma_5^-$ (polar distortion) in $A2_1am$ phase of LBSI-$x$ in the Sr substitution range of 0-0.9 at room temperature.

Next, we clarify the impacts of disordered A-cations distribution on $a^0a^0c^+$ OOR distortion. Since our computational framework does not allow fractional occupancies, it is necessary to examine the crystal structures with many different specific distributions of Sr/La to enable comparisons between theory and experiment. A series of structures with differing A-cations distributions of Sr/La for LSIO and Ba/La for LBIO were systematically generated, starting from the A-site ordered $P4_2/mnm$, $Pnam$, and $A2_1am$ supercells within a range of four formula units (48 atoms). Hereafter, we refer to the A-site-disordered structures derived from three most stable structures as their supercells, respectively, although the generated disordered structures no longer belong to their space group. As a result, 49, 77 and 453 polymorphs with varying distributions of A-cations for LBIO and LSIO were produced from the $P4_2/mnm$, $Pnam$, and $A2_1am$ supercells, respectively, to conduct further structural relaxation. The generated structures are classified into the following $A^{2+}/La^{3+}$ ratios inside the perovskite slab $[A/La]_{Pv}$ (rocksalt layer $[A/La]_{Rs}$): 100/0 (0/100), 75/25 (12.5/87.5), 50/50 (25/75), 25/75 (37.5/62.5), and 0/100 (50/50) corresponding to the

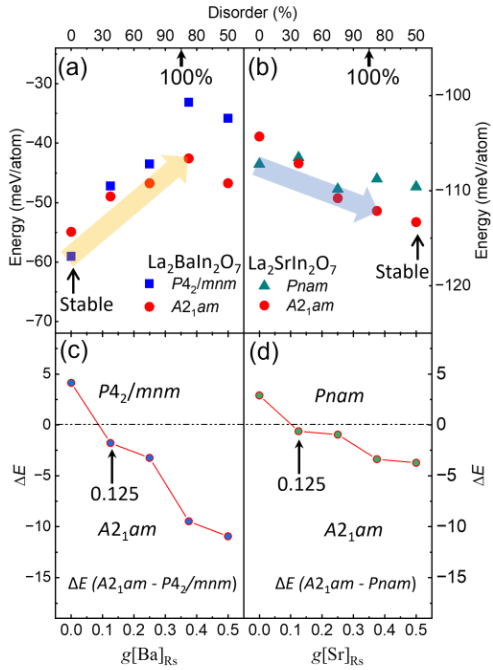

Figure 8. (a) two structures lowest relaxed energy of La$_2$BaIn$_2$O$_7$. the difference in the lowest relaxed energy between $A2_1am$ and $P4_2/mnm$, $\Delta E$, against varied Ba/La distribution on the A$_{Rs}$-site (occupancy of Ba on the A$_{Rs}$-site, $g[Ba]_{Rs}$) for La$_2$BaIn$_2$O$_7$ and the difference in the lowest relaxed energy between between $A2_1am$ and $Pnam$, $\Delta E$, against varied Sr/La distribution on the A$_{Rs}$-site (occupancy of Sr on the A$_{Rs}$-site, $g[Sr]_{Rs}$) for La$_2$SrIn$_2$O$_7$.

occupancy of $A^{2+}$ cation on the A$_{Rs}$-site $g[A^{2+}]_{Rs}$ = 0.0, 0.125, 0.25, 0.375, and 0.5, respectively. Using the same procedure as previously reported for La$_2$SrSc$_2$O$_7$,[61] we explored structural trends across a range of different cation distributions.

The examination of the relaxed structures (see Figure S6, SI) reveals that the A-cations distribution has an impact on the energy stability of the investigated supercells, as shown in Figure 8a and 8b. The energy difference between the two states with lowest total energies, $\Delta E_{LBIO} = E_{A2_1am} - E_{P4_2/mnm}$ for La$_2$BaIn$_2$O$_7$ and $\Delta E_{LSIO} = E_{A2_1am} - E_{Pnam}$ for La$_2$SrIn$_2$O$_7$, as a function with each $g[A^{2+}]_{Rs}$ are investigated theoretically (see Figure 8c and 8d). To conveniently parameterize the magnitude of A-site cation disorder (in percentage), we define the order factor $3 \cdot |g[A^{2+}]_{Rs} - 1/3| \times 100\%$ as a function of $g[A^{2+}]_{Rs}$. The disorder factor (σ) of A-cations is thus defined by a function of σ = 1 - $3 \cdot |g[A^{2+}]_{Rs} - 1/3| \times 100\%$. The disorder σ = 0% of A-cations is assigned to the ordered distribution of A-cations, $g[A^{2+}]_{Rs} = 0$ (i.e., the A$_{Rs}$-sites are fully occupied by La$^{3+}$, $g[La^{3+}]_{Rs} = 1.0$), and the disorder factor σ = 100% is assigned to the fully disordered distribution of A-cations, $g[A^{2+}]_{Rs} = 1/3$ (i.e., $g[A^{2+}]/g[La] = 0.33/0.67$ at each A-site). Therefore, the disorder σ = 50 % is assigned as the A$_{Pv}$-site with full occupancy of La$^{3+}$ ($g[A^{2+}]_{Rs} = 0.5$), rather than 0% or 100% disorder, due to the disordered arrangement of A-cations on the A$_{Rs}$ sites ($[A^{2+}/La^{3+}]_{Rs} = 50/50$). 100% disorder is defined as the $[A^{2+}/La^{3+}]_{Pv} = [A^{2+}/La^{3+}]_{Rs} = 33/67$.

For LBIO shown in Figure 8c, it is observed that the polar $A2_1am$ supercell is more stable than nonpolar $P4_2/mnm$ structure, when the $g[Ba]_{Rs} \geq 0.125$ (σ ≥ 37.5%). However, for all the structures explored, the most stable structure is $P4_2/mnm$ with ordered A-cations (Ba and La), and the energy rises with high disorder of A-cations, indicating that it is energetical favorable for all Ba to occupy the A$_{Pv}$-site only. This result is consistent with the experimentally observed absence of Ba at A$_{Rs}$-site (Figure 3) for all investigated LBSI-$x$ series. In contrast, for LSIO system, the energy gradually descends with high disorder of A-cations for all explored structures (Figure 8b). For LSIO, the energy of explored structure descends with $g[Sr]_{Rs}$ indicating that the introduction of Sr in rocksalt layer is energetical favorable. When $\Delta E_{LSIO} < 0$ at $g[Sr]_{Rs} \geq 0.125$, the polar $A2_1am$ supercell becomes more stable than $Pnam$ that is the stable structure as A$_{Pv}$-sites are fully occupied by Sr ($g[Sr]_{Rs} = 0$). It is obvious that the most stable $A2_1am$ supercell of LSIO with absent Sr in perovskite slab ($g[Sr]_{Rs} = 0.5$) is energetically superior to all explored structures (Figure 8b). Therefore, the substitution of Sr for Ba in LBSI-$x$ series rises the disordered distribution of A-cations (Sr and La) resulting in the stabilized HIF $A2_1am$ structure by suppressing the competition with nonpolar $P4_2/mnm$ and $Pnam$ structures.

The realization of new HIF La$_2$(Ba,Sr)In$_2$O$_7$ with A-cations disorder offers a better chance to develop potential multiferroic/magnetoelectric materials from $n = 2$ RP family of $Ln^{3+}_2A^{2+}B_2O_7$, because the presence of abundant trivalent magnetic cations suitable for octahedral coordination on the B-site enables the rational design of a large class of such materials. In the present case, the large B-site cation such as Sc$^{3+}$ is introduced to stabilize the highly distorted perovskite frameworks required for the HIF mechanism to work, but it will be possible to induce HIF for the smaller B-site cations (*e.g.*, 3d transition metal cations) by reducing the size of $Ln^{3+}$ and $A^{2+}$ cations while keeping the small size difference between these two cations. From the synthesis perspective, the high-temperature solid-state reactions required for the preparation of most complex metal oxides entropically tend to form cation-disordered phases. Even if such phases are not synthesized by the conventional solid-state reaction, nonequilibrium synthesis methods like solution-based processing and vapor-phase deposition are expected to stabilize the cation-disordered HIF phases.

## 4. CONCLUSION

In this work, the series of In-based layered perovskites $La_2Ba_{1-x}Sr_xIn_2O_7$ for $0.0 \leq x \leq 0.9$ in RP phase are synthesized by complex polymerization (so-gel) method and an intriguing temperature-composition ($T$ - $x$) phase diagram has been obtained. The role played by the chemical substitution-induced two non-polar structure distortions in stabilizing ferroelectricity is demonstrated by combination of experimental results and first-principles calculations. Specifically, our structural analyses and ferroelectricity studies verify the HIF $n$ = 2 RP solid solutions for $0.5 \leq x \leq 0.9$ in $A2_1am$ phase at room temperature and these solid solutions undergo a second-order phase transition from ferroelectric to paraelectric upon elevation of temperature. For small $x$ ($\leq$ 0.3), on the other hand, the remarkable competition of the interlayer rumpling mainly drives the disappearance of oxygen octahedral rotation resulting in the condensed paraelectric $P4_2/mnm$ phases at room temperature and these solid solutions undergo a first-order phase transition of $P4_2/mnm$-to-$Amam$ at high temperature. The experimental results, supported by first-principles calculations, clarify the role of A-site cation distribution in restoring the OOR and stabilizing the ferroelectric structure. Disordering of $Sr^{2+}/La^{3+}$ cations at A-sites in perovskite slabs suppresses the rumpling-induced octahedral deformation and instead facilitates the $InO_6$ octahedral rotations. Eventually, the suppression of interlayer rumpling, in addition to the small tolerance factor, allows the concurrence of the OOR and OOT distortions, which stabilizes the hybrid-improper ferroelectricity. Our findings demonstrate that the interplay of the disordered A-site cations Sr/La and the $BO_6$ octahedral rotation will be exploited as a knob to explore a new class of ferroelectric materials in $n$ = 2 RP-type perovskites.

## 5. METHODS

**Synthesis**

Polycrystalline samples of $n$ = 2 RP-type $La_2Ba_{1-x}Sr_xIn_2O_7$ ($x$ = 0, 0.1, 0.2, 0.3 0.4 0.5 0.6 0.7 0.8 and 0.9) were synthesized by using polymerizable complex methods. Reagent-grade $SrCO_3$ (99.9%, Kojundo Chemical Laboratory Co., Ltd.), $BaCO_3$ (99.9%, Kojundo Chemical Laboratory Co., Ltd.), $La_2O_3$ (99.99%, Kojundo Chemical Laboratory Co., Ltd., dried at 900 °C), and $In_2O_3$ (99.99%, Kojundo Chemical Laboratory Co., Ltd.) powders used as the starting materials were successively dissolved in nitric acid according to the stoichiometric ratio. $La_2O_3$, $BaCO_3$, $SrCO_3$, $In_2O_3$ and citric acid were dissolved by heating at 250 °C for 2h to obtain a transparent solution. After ethylene glycol was added, the mixture was heated at 200 °C over 6 h under stirring to promote polymerization. The resultant gel was further heated at 400 °C in air for 6 h. The precursor powders thus obtained were ground, pressed into pellets, and calcined at 1000 °C in air for 12 h to remove carbon residue. The calcined powders were reground and pelletized again and slowly heated up to 1300~1540 °C in air and kept for ~24 h, followed by naturally cooling (150 °C/h) to room temperature.

**Characterization**

The phase purity of samples was assessed using X-ray powder diffraction (XRD) data collected by a laboratory X-ray diffractometer (SmartLab, Rigaku) using the monochromatic Cu $K_α$ radiation. High-resolution SXRD was taken at variable temperature range from 300 K to 1000 K using a large Debye-Scherrer camera with MYTHEN solid-state detectors installed at SPring-8 BL02B2.[65] The monochromatic X-ray wavelengths of $λ$ = 0.7747021 Å were used. The finely ground powder samples were housed in capillaries, which were rotated continuously during measurement to diminish the effect of preferred orientation. The Lindeman capillary with 0.1 mm inner diameter and the $SiO_2$ capillary with 0.2 mm inner diameter were used for the measurements at room temperature and high temperature, respectively. High-resolution time-of-flight (TOF) NPD at room temperature was performed using the SuperHRPD diffractometer at J-PARC.[66, 67] The powder samples were packed into a vanadium can with 6 mm inside diameter. Using the backscattering, 90º, and 30º bank detectors, diffraction data were collected over a TOF range of 20-130 ms, corresponding to a $d$-range of 0.41-16.0 Å. Based on these diffraction data, the whole-pattern decompositions were performed to test the space groups by using the Le Bail method[68] and the crystalline structures were analyzed by the Rietveld refinements [69] with the JANA2006.[70] The combined refinements against the NPD and SXRD data obtained at the same temperatures were carried out by the Rietveld method with the JANA2006. Symmetry analysis and distortion-mode quantification were performed using the web-based tool ISODISTORT program.[71]

Optical SHG measurements were performed on polycrystalline pellets with a 1064 nm beam emitted from Nd: YAG laser with 100 ps pulses and 10 Hz repetition rate. The measurements were carried out in reflection geometry with a 532 ± 10 nm bandpass filter. Temperature-variable experiments were carried out using an electric heater for heating and cooling processes. Electric polarization versus electric field (*P-E*) hysteresis loops were measured with the positive-up negative-down (PUND) method.[72] The chemical composition and atomic distribution of cations were examined by EDS using a Super-X detector attached to STEM. The HAADF-STEM images were acquired on a 300 keV Thermo Scientific Titan 60-300 with a double corrector set.

**Theoretical Calculations**

First-principles density functional theory (DFT) calculations were carried out using the projector augmented-wave (PAW) method as implemented in the Vienna *ab-initio* Simulation Package (VASP) within the generalized gradient approximation (GGA).[73-76] The Perdew-Berke-Ernzerhof exchange-correlation functional revised for solids (PBEsol) was used throughout this work.[77] A cutoff energy for plane waves was set to 550 eV, and the standard PAW radial cutoffs were used. The following electronic states were treated as valence electrons: $2s^2$ and $2p^4$ for O; $3d^1$ and $4s^2$ for Sc; $4s^2$, $4p^6$, and $5s^2$ for Sr; $5s^2$, $5p^6$, and $6s^2$ for Ba; and $5s^2$, $5p^6$, $5d^1$, and $6s^2$ for La. $k$-space sampling meshes were generated with an interval of $2\pi \times 0.04$ Å$^{-1}$ using a Γ-centered scheme. Lattice constants and internal coordinates were relaxed until residual stress and forces became less than 0.01 GPa and 1 meV Å$^{-1}$, respectively. Phonon frequencies were derived from the calculated force constants using the PHONOPY code.[78] A-cation arrangements were exhaustively searched for by using the CLUPAN code.[79]

**ASSOCIATED CONTENT**

**Supporting Information**

Supporting Information is available from the Wiley Online Library or from the author.

**AUTHOR INFORMATION**

**Corresponding Author**

*yi.wei.5c@kyoto-u.ac.jp

*fujita.koji.5w@kyoto-u.ac.jp

**ACKNOWLEDGMENT**

This work was supported by the Grant-in-Aid for the JSPS KAKENHI (Grant Numbers JP22K04686 JP19H02433, JP20K20546, JP21H04619, JP22F21750, and JP22H01775), the JSPS Research Fellow (Grant Number 22KF0195), and the "Advanced Research Infrastructure for Materials and Nanotechnology in Japan (ARIM)" of the MEXT (Proposal Number JPMXP1222HK0061). SXRD experiments were performed on BL02B2 at SPring-8 with the approval of JASRI (Proposal Nos. 2022A1587, 2022B0553, and 2022B1668). Time-of-flight ND experiments on SuperHRPD at the J-PARC Pulsed Neutron and Muon Source were supported by beam time allocations from STFC (Proposal No. 2022A0067).